\documentclass[showpacs,amsmath,amssymb,floatfix,prd,superscriptaddress,preprint]{revtex4-1}
\usepackage{amssymb,scalerel}
\usepackage{graphics}
\usepackage[T1]{fontenc}
\usepackage[utf8]{inputenc}
\usepackage{epstopdf}
\usepackage{epsfig}
\usepackage{dcolumn}
\usepackage{bm}
\usepackage{mathtools}
\renewcommand{\vec}[1]{\mathbf{#1}}
\newcommand{\gv}[1]{\ensuremath{\mbox{\boldmath$ #1 $}}} 
\newcommand{\abs}[1]{\left| #1 \right|} 
\newcommand{\avg}[1]{\left< #1 \right>} 
\newcommand{\pd}[2]{\frac{\partial #1}{\partial #2}} 
 
\let\baraccent=\= 
\renewcommand{\=}[1]{\stackrel{#1}{=}} 
 
\usepackage{color}
 \definecolor{blue}{rgb}{0,0,1} 
 \definecolor{sepia}{rgb}{0,0.8,0.2}
 \definecolor{redi}{rgb}{0.5176,0.0078,0.0078}

\begin{document}

\title{Collective self-optimization of communicating active particles}



\author{Alexandra V. Zampetaki}
\affiliation{Max-Planck-Institut f\"{u}r Extraterrestrische Physik, 85741 Garching, Germany}
\affiliation{Institut f\"{u}r Theoretische Physik II, Weiche Materie, Heinrich-Heine-Universit\"{a}t, 40225 D\"{u}sseldorf, Germany}

\author{Benno Liebchen}
\email[]{benno.liebchen@pkm.tu-darmstadt.de}
\affiliation{Institute of Condensed Matter Physics, Technische Universit\"{a}t Darmstadt, 64289 Darmstadt, Germany}

\author{Alexei V. Ivlev}
\affiliation{Max-Planck-Institut f\"{u}r Extraterrestrische Physik, 85741 Garching, Germany}

\author{Hartmut L\"{o}wen}
\affiliation{Institut f\"{u}r Theoretische Physik II, Weiche Materie, Heinrich-Heine-Universit\"{a}t, 40225 D\"{u}sseldorf, Germany}

\date{\today}

\begin{abstract}
The quest on how to collectively self-organize in order to maximize
the survival chances of the members of a social group requires finding 
an optimal compromise between maximizing the well-being of an individual
and that of the group. Here we develop a minimal model describing active
individuals which consume or produce, and respond to a shared resource, 
– such as the oxygen concentration for aerotactic bacteria or the temperature
field for penguins – while urging for an optimal resource value. Notably, 
this model can be approximated by an attraction-repulsion model,
but in general it features many-body interactions. While the former prevents 
some individuals from closely approaching the optimal value of the shared 
"resource field", the collective many-body interactions induce aperiodic patterns, 
allowing the group to collectively self-optimize. Arguably, the proposed
optimal-field-based collective interactions represent a generic concept at
the interface of active matter physics, collective behavior, and microbiological
chemotaxis. This concept might serve as a useful ingredient to optimize ensembles of synthetic
active agents or to help unveiling aspects of the communication rules which 
certain social groups use to maximize their survival chances. 
\end{abstract}

\maketitle
\begin{center}
	\textbf{I. INTRODUCTION}
\end{center}

The survival of organisms hinges to a great extent on their adaptation ability to environmental changes. One of the simplest, yet effective, adaptation mechanisms is provided by chemotaxis \cite{Wadhams2004, Eisenbach2007}. Many microorganisms can sense the concentration of a chemical and move  up
(positive chemotaxis) or down (negative chemotaxis) its gradient \cite{Adler1969, Macnab1972}. This ability does not only allow them to navigate towards nutrients and away from dangerous toxins, but it can also be used for inter-cell signalling
\cite{Speranza2010, Friedrich2008, Hoell2011,Kaupp2008, Grima2005, Sengupta2009, Kollmann2005}. In particular,
{\it E. Coli} bacteria and {\it Dictyostelium} cells, when starving, produce collectively  certain chemicals \cite{Berg1972,Bonner1947,Gregor2010,Veltman2008,Reid2016}. Each of these microorganisms then follows the collective chemical  gradient, yielding their aggregation, which enables them to endure long starvation periods.
This chemotactic auto-aggregation is well established \cite{Murray2003,Painter2019,Marsden2014, Liebchen2019,Stark2018,Chertock2019} and captured  within the classic Patlak-Keller-Segel  model \cite{Keller1970,Keller1971}. 
More recent variants of this model \cite{Hillen2009}, such as the volume-filling model, prevent the occurrence of infinitely dense aggregates \cite{Hillen2001a,Wrzosek2006,Ma2012} by limiting the density of growing aggregates, as relevant for cellular systems  \cite{Painter2002,Wrzosek2010,Wang2007,Bubba2020}. In this respect, finite volume effects are taken into account, by considering that the occupation of a certain area by finite-volume cells restricts the vacant space and thus hinders other cells from  inhabiting it.

The availability of space in the environment is not the only decisive factor that can regulate the migration of organisms.
Nature is also full of cases where both an excess and a deficit 
in a  specific ambient quantity can be harmful for the considered organism. 
One example are aerotactic bacteria, whose motions are dictated by the urge for an  optimal oxygen concentration \cite{Taylor1983,Taylor1999,Zhulin1996,Mazzag2003,Elmas2019}.
Furthermore, several phototactic algae, such as  {\it Euglena gracilis}, are known to seek optimal levels of light intensity for photosynthesis \cite{Giometto2015}.
For more complex living systems,  the maintenance of a state of  optimal conditions - homeostasis- can be achieved both by physiological  and  behavioral responses \cite{Schulkin2004}. This introduces a paradigm 
of  {\it social} thermoregulation   for different species, such as birds \cite{Douglas2017}, mice and rats \cite{Alberts1978,Gordon1990,Glancy2015}, with the most remarkable being the case of penguins \cite{Ancel1997,Gilbert2008,Ancel2015,Gilbert2006,Gerum2018}. In cold conditions, such animals aggregate into huddles-  which they break apart, once their upper comfort temperature is surpassed, rendering the huddling formation a highly dynamical process.

In all the aforementioned cases, the accomplishment of a nearly optimal state for each individual organism is astonishing and 
its feasibility seems to rely on their   communication and co-operation. A primitive communication scheme, established within the framework of chemotaxis,  consists in the signalling via the  collective production and response to certain chemicals. Typical examples thereof constitute the production of cAMP from {\it Dictyostelium} cells \cite{Veltman2008, Reid2016}, pheromones from ants \cite{Wilson1971,Sumpter2003,Jackson2006}, and autoinducer 2 by {\it E. Coli} bacteria  \cite{Laganenka2016}, which allow the individuals to exchange information and regulate their motion towards a certain goal.  For a collective goal of enhanced difficulty, such as to achieve an optimal state for all the members of a colony, it is questionable whether 
this simple chemotactic strategy can guarantee its realization.



\begin{figure}[htbp]
	\begin{center}
		\includegraphics[width=0.5\columnwidth]{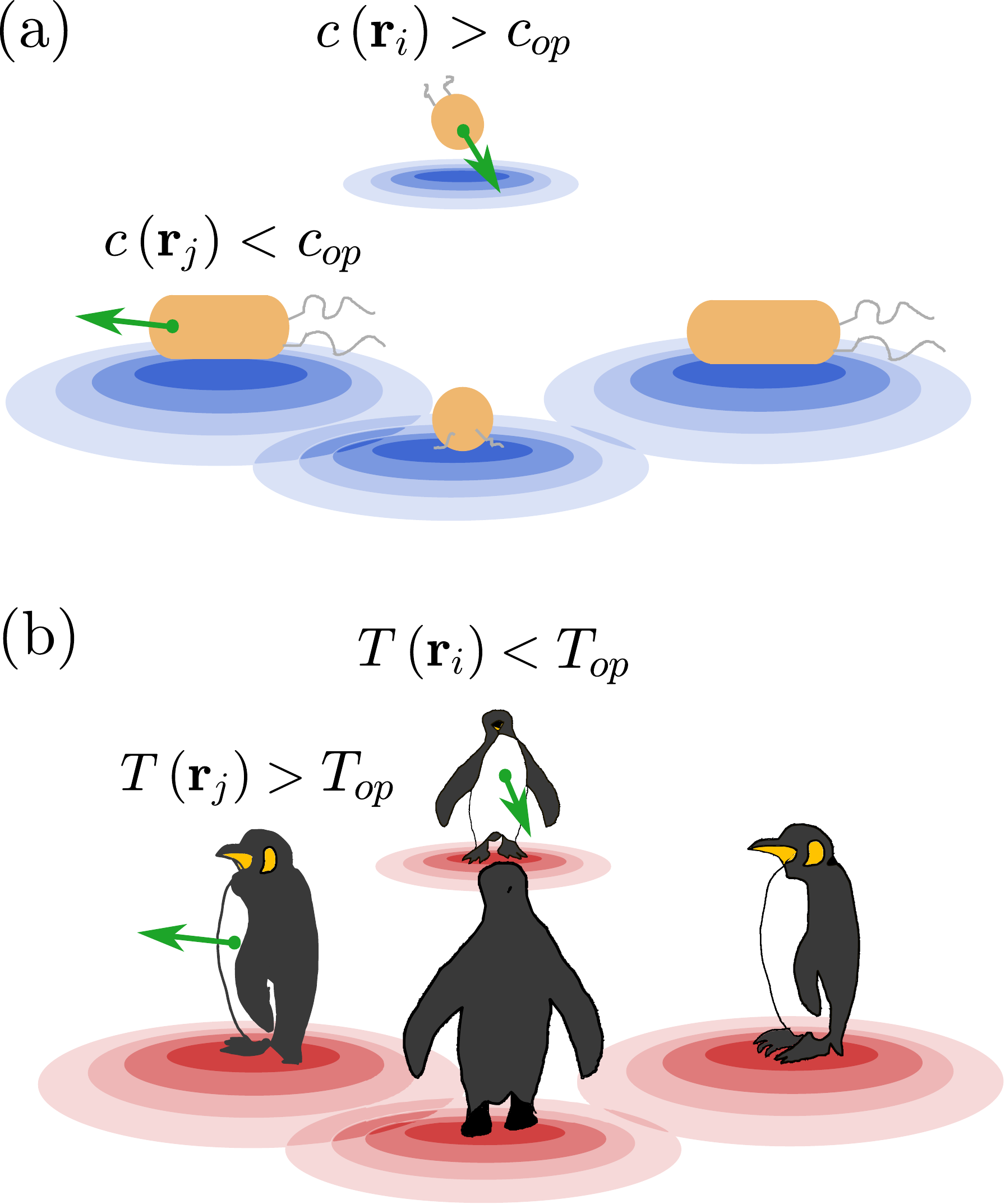}
	\end{center}
	\caption{\label{pen_fig} Schematic illustration of two different systems to which our model is applicable: (a) aerotactic bacteria with an optimal oxygen concentration $c_{op}$, acting as oxygen sinks (blue color) and (b) social thermoregulation in emperor penguins with a comfortable temperature $T_{op}$, acting as heat sources (red color) at their positions $\vec{r}_k$. The green arrows in the subfigures point to the direction in which the respective bacteria (a) or penguins (b) tend to move (away or towards the other group members) in order to optimize the field value in their individual positions.   }
\end{figure}

In the present work we introduce a generic physical model allowing us to explore the effectiveness of a straightforward chemotaxis-inspired strategy for the collective optimization of individuals.
In particular, we consider agents who, similarly to chemotactic bacteria, produce or consume a certain scalar field  (e.g. oxygen concentration for aerotactic bacteria or  the temperature field for penguins, Fig. \ref{pen_fig}), and  move up or down its gradient, in order to approach their individual optimum (e.g. optimal oxygen concentration or optimal temperature).  
Our model is distinct from the volume-filling model for chemotaxis \cite{Painter2002}, where the existence of an optimal density, instead of an optimal field, determines the system's evolution.
We demonstrate  that the proposed  communication rules lead to the collective self-optimization of the system, allowing nearly all individuals to reach an optimal field state.

More specifically, the very existence of an optimal field value causes our model to reduce to an attraction-repulsion model in the adiabatic limit where the dynamics of the scalar field is fast. Such models feature in general an optimal interparticle  distance,  which promotes the formation of closely packed structures \cite{Phillips1981,Stillinger2006,Chuang2007}. Here, however, the inherent nonlinearity of the model, in terms of the  scalar field, 
creates effective three-body interactions, resulting in the  formation of aperiodic clusters with a high accumulation of individuals on the periphery. These clusters turn out to belong to a smooth manifold of equilibria where most individuals are near the optimal state. Interestingly,
the presence of a weak noise allows the individuals to explore this manifold and thus provides them with  numerous pathways to reach 
the highly degenerate optimal state.  The underlying dynamics is rather intricate and characterized by varying spatial structures, whose description lies beyond standard mean-field models. We envisage that the present work could serve as a
starting point for generating new insights
at the interface of biology, social-systems-study and physics through
quantitative modelling.

 \begin{center}
 	{\textbf{II. MODEL}}
 \end{center}
 \begin{center}
	{\textbf{A. Model Definition}}
\end{center}
Our goal here  is to model situations where a number $N$ of active particles  try to self-optimize their positions $\{r_k\}$ in respect to an optimal 
value $u_{op}$ of the collective scalar field they self-produce or consume. Examples thereof represent the conformations of aerotactic bacteria, consuming the ambient oxygen while striving for an optimal oxygen concentration $c_{op}$ (Fig. \ref{pen_fig} (a)), or of social thermoregulation e.g. in mice, birds or penguins, acting as heat sources and seeking an optimal 
temperature $T_{op}$ (Fig. \ref{pen_fig} (b)). For simplicity we discuss the case where each particle is a point source of a scalar field $u(\vec{r},t)$ at its position $\vec{r}_i$ at time $t$ \footnote{The connection to the case where agents acts as sinks (Fig. \ref{pen_fig} (a)) is straightforward}.

We assume that the evolution of this field is governed by the following  three dimensional (3D) diffusion equation, regarding e.g.  diffusion in a solvent or heat induction,
\begin{equation}
\pd{u}{t}=D \nabla^2  u-k_d u +k_s \sum_j \delta \left(\vec{r}-\vec{r}_j\right),\label{dif_eq}
\end{equation}
where  $D$ denotes the field diffusion constant, $k_s$ represents the emission rate of each particle source and the sum runs over all
particles indices $j$. The constant $k_d$ represents the loss rate due to possible external factors, such as secondary chemical reactions or wind advection. We remark that $k_d=0$ serves  as an important special case of  Eq. \ref{dif_eq}, regarding field diffusion  in the absence of any external losses.

The steady state solution for the collective field $u$ can be expressed as a superposition of single particle Yukawa orbitals \cite{Liebchen2019} as follows
\begin{equation}
u(\vec{r})=\sum_{j}Y\left(\left| \vec{r}-\vec{r}_j\right|\right)= \sum_{j} A\frac{\exp\left(-\kappa\left| \vec{r}-\vec{r}_j\right|\right)}{\left| \vec{r}-\vec{r}_j\right|} . \label{eq_temp}
\end{equation}
Here we have introduced the parameters $\kappa=\sqrt{k_d/D}$ and $A=k_s/(4\pi D)$ and the sum runs over all particles $j$.  Notably, due to their Yukawa character, these orbitals strongly resemble to the potential function of charged colloids, suggesting a  connection between the two systems (see also SI). 

In order to proceed to a Lagrangian formulation of our model on the level of individual particles, we make  the additional assumption that the scalar field $u(\vec{r},t)$ equilibrates much faster than the timescale  in which the particles move, which is very well justified e.g. in the case of heat production for penguins or chemical production by microorganisms. Then the field sensed by the particle $i$ in its position $\vec{r}_i$ can be  expressed as $u(\vec{r}_i)=\sum_{j\neq i}Y\left(\left| \vec{r}_i-\vec{r}_j\right|\right)$, where the self-interaction is ignored.  We note that a similar assumption
of fast chemoattractant dynamics has been employed to express the original  Patlak-Keller-Segel model in terms of a logarithmic interaction kernel \cite{Fellner2011}, accounting for non-local attraction forces, which cause the system's aggregation \cite{Fellner2011,Topaz2006,Milewski2008}.

Here, we assume  that each particle $i$, located
at a position $\vec{r}_i$, moves in the direction of $\nabla_i u(\vec{r}_i)$ if $u(\vec{r}_i)<u_{op}$ and in the opposite direction if $u(\vec{r}_i)>u_{op}$, with $u_{op}$ the optimal field value. This leads to the following  effective force  $\vec{F}_i$, governing the motion of the $i$-th particle 
\begin{equation}
\vec{F}_i=-\lambda\left(u(\vec{r}_i)-u_{op}\right)\nabla_i u(\vec{r}_i), \label{full1}
\end{equation}
with  $\lambda>0$  being the alignment strength along 
the field gradient. 
In the following we use dimensionless units, by measuring the length, the field and the time  in units of $x_u=\kappa^{-1}$, $u_u= A \kappa$ and $t_u=\gamma/(\lambda A \kappa^4)$ respectively, with $\gamma$ denoting the Stokes drag coefficient. We also assume that the particles are allowed to move on a two dimensional (2D) plane. 

We note that, on the mean-field level, the model described here resembles the volume-filling model for chemotaxis \cite{Hillen2001a,Painter2002,Wrzosek2010}. In particular, our model assumes a field-dependent chemotactic sensitivity, guiding the system towards an optimal field value $u_{op}$, whereas the volume-filling model  features a density-dependent sensitivity that directs the system towards an optimal density $\rho_{op}$. Interestingly, while
both models exhibit the same linear behaviour,
and feature a transition from a uniform to a patterned state for field/density values lower than the optimal ones, they significantly differ in the nonlinear regime in a way which is crucial to allow all individuals in the system to closely approach the optimal field value. A more detailed comparison between the two models is presented in the supplementary information (SI).   

\begin{center}
	{\textbf{B. Three-body interactions}}
\end{center}

Taking into account Eqs. \ref{eq_temp}, \ref{full1} and introducing $r_{i,j}=\left| \vec{r}_i-\vec{r}_j\right|$, the effective force $\vec{F}_i$ can be rewritten as $\vec{F}_i=\vec{F}_{i,p}+\vec{F}_{i,t}$, where the part
\begin{equation}
\vec{F}_{i,p}=-\frac{1}{2}\nabla_i\sum_{j\neq i}\left[Y\left(r_{i,j}\right)-u_{op}\right]^2 \label{pair_int}
\end{equation}
consists only of pair interactions between the particle $i$  and any other particle $j$. This part is of Hamiltonian nature, since $\vec{F}_{i,p}=-\nabla_i V$,
with $V=\frac{1}{4}\sum_k \sum_{j\neq k} \left[Y\left(r_{k,j}\right)-u_{op}\right]^2$ being the total  interaction potential.  Given that the Yukawa orbitals $Y\left(r_{k,j}\right)$
are decreasing functions of the interparticle distances $r_{k,j}$, this potential sets essentially an optimal interparticle distance in the system. On a qualitative level, this resembles  the requirement of an optimal density, allowing us to draw a connection between this Hamiltonian part and the aforementioned volume-filling model.

\begin{figure}[htbp]
	\begin{center}
		\includegraphics[width=0.7\columnwidth]{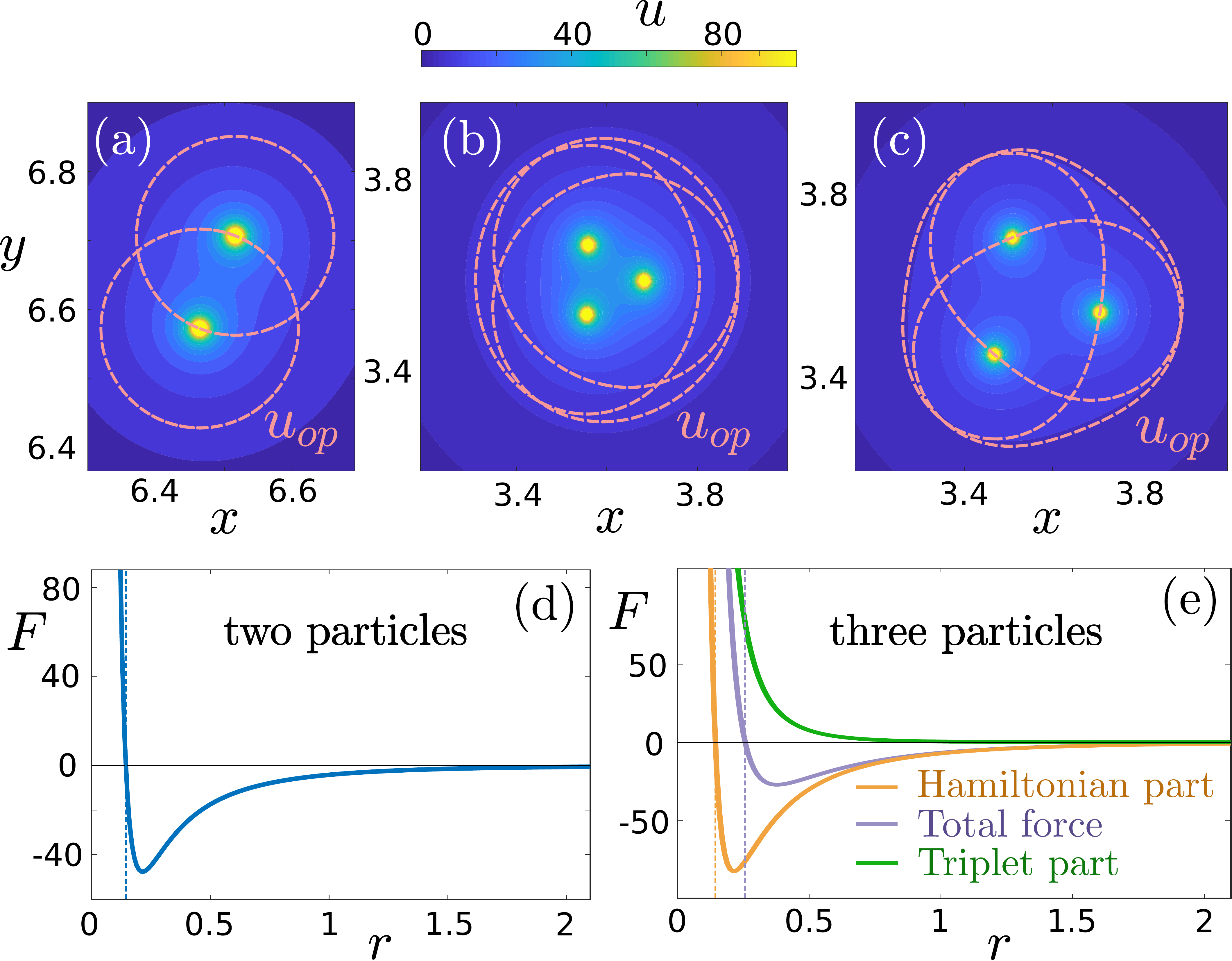}
	\end{center}
	\caption{\label{tri_cha1}Three-body interactions allow all individuals to reach the optimal state: (a)-(c) Contour plot of the collective field ${u}(x,y)$ (Eq. \ref{eq_temp})
		for the equilibrium positions $\{\vec{r}_i^{(0)}\}$ of (a) two particles, (b) three particles in the Hamiltonian approximation
		$\vec{F}_i\approx\vec{F}_{i,p}$ and (c) three particles in the non-Hamiltonian model (Eq. \ref{full1}). Each dashed pink line is assigned to a different particle $i$ and depicts the set of points $(x_i,y_i)$ which satisfy the condition $u\left(\vec{r}_i\right)=u_{op}$ for this particle. (d) The  force projection $F=\vec{F}_i\cdot \hat{\vec{r}}_{i,j}$  for a two-body system as a function of the interparticle separation $r$. 
		For three particles the only candidate equilibrium is that of an equilateral triangle with a side $r$. The force projection $F$  for this case as a function of the side length $r$  is shown in subfigure (e) where the total force projection (purple), its Hamiltonian part (orange) and its triplet part (green) are shown. In all cases $u_{op}=6$ and the vertical lines mark the respective equilibrium distances $r^{(0)}$. The considered particles  move in a 2D plane.}
\end{figure}

In contrast, the part
\begin{equation}
\vec{F}_{i,t}=-\nabla_i\left[\sum_{j\neq i} \sum_{l\neq i, l>j}Y\left(r_{i,j}\right)Y\left(r_{i,l}\right)\right] \label{trip_int}
\end{equation}
consists of triplet interactions between particles $i$,$j$ and $l$. Contrary to other forms of triplet interactions \cite{Lowen1998,Russ2002}, this term cannot be associated to any total potential function. Thus, the corresponding system  possesses a \textit{non-Hamiltonian} character, implying that it cannot be described by standard
statistical equilibrium mechanics \cite{Ivlev2015}.   Evidently, such a term would affect also the mean-field behaviour of the model, giving rise to a triplet rather than a pairwise sensing kernel \cite{Fellner2011}, which cannot be described in terms of a free energy functional.

The effect of the peculiar triplet interaction term $\vec{F}_{i,t}$, can be observed in Fig. \ref{tri_cha1}. In the two-particle case, $\vec{F}_{i,t}=0$  and the system is Hamiltonian. Then the projection $F$ of the  force $\vec{F}_i$ on the direction of $\vec{r}_{i,j}$, $F=\vec{F}_i\cdot \vec{r}_{i,j}/\abs{\vec{r}_{i,j}}$, exhibits a Lennard-Jones-like character with a repulsive and an attractive part (Fig. \ref{tri_cha1} (d)) and a single equilibrium  distance $r^{(0)}$, where each  particle lies at its optimal field value $u_{op}$ (Fig. \ref{tri_cha1} (a)).   This two-body behavior is 
already different from that of the Patlak-Keller-Segel model for chemotaxis  (featuring only attractive non-local interactions) and resembles instead nonlocal repulsion-attraction models, such as the D'Orsogna model \cite{Orsogna2006,Chuang2007,Leverentz2009}, used often to explore  distinctive properties of collective behavior.

For three particles,
triplet interactions start to play an important role. In the Hamiltonian approximation, in which these 
are neglected and $\vec{F}_i\approx\vec{F}_{i,p}$, the forces  (Eq. \ref{pair_int}) cannot correctly capture the consequences of assuming that an optimal field value $u_{op}$ exists.
In particular, while in both the Hamiltonian approximation (Eq. \ref{pair_int}) and the full non-Hamiltonian  model (Eq. \ref{full1}) the particles equilibrate in an equilateral triangle structure (Fig. \ref{tri_cha1} (b),(c)), only in the latter case they manage to fulfil their optimal condition, $u(\vec{r}_i)=u_{op}$ (Fig. \ref{tri_cha1} (c)).
In the absence of the  strictly  repulsive triplet forces $\vec{F}_{i,t}$,
the particles equilibrate  instead at a closer distance, dictated by the Hamiltonian pairwise interactions $\vec{F}_{i,p}$ (Fig. \ref{tri_cha1} (e)). In  this case it follows that $u(\vec{r}_i)>u_{op}$ (Fig. \ref{tri_cha1} (b)), highlighting the fact that the three-body interactions constitute an integral part of  our model.

 \begin{center}
	{\textbf{III. RESULTS}}
\end{center}

\begin{center}
	{\textbf{A. Phase diagram}}
\end{center}

As shown in Fig. \ref{tri_cha1} (e), the interactions in the system, even with the inclusion of the triplet non-Hamiltonian term, possess an attractive and a repulsive part. Therefore, on the many-particle level, our system is expected  to display  overall a similar  phase behavior to that of Lennard-Jones   or Morse potential \cite{Orsogna2006,Chuang2007} systems, however with important deviations in the local structures that emerge. In particular, from the form of the interaction force we expect  a solid (uniform) phase for high number densities $\rho>\rho^*$, for which the optimal equilibration is prohibited, and a solid-vacuum  coexistence (patterned phase) for small densities $\rho<\rho^*\propto (\bar{d}^*)^{-1/2}$, where the particles form clusters with an optimal mean inter-particle separation $\bar{d}^*$ \cite{Phillips1981, Stillinger2006}. Interestingly, this solid-vacuum coexistence resembles huddling in penguin colonies, forming dense aggregates when the outside temperature 
decreases below a certain threshold \cite{Ancel1997,Gilbert2008,Ancel2015}.

\begin{figure}[htbp]
	\begin{center}
		\includegraphics[width=0.7\columnwidth]{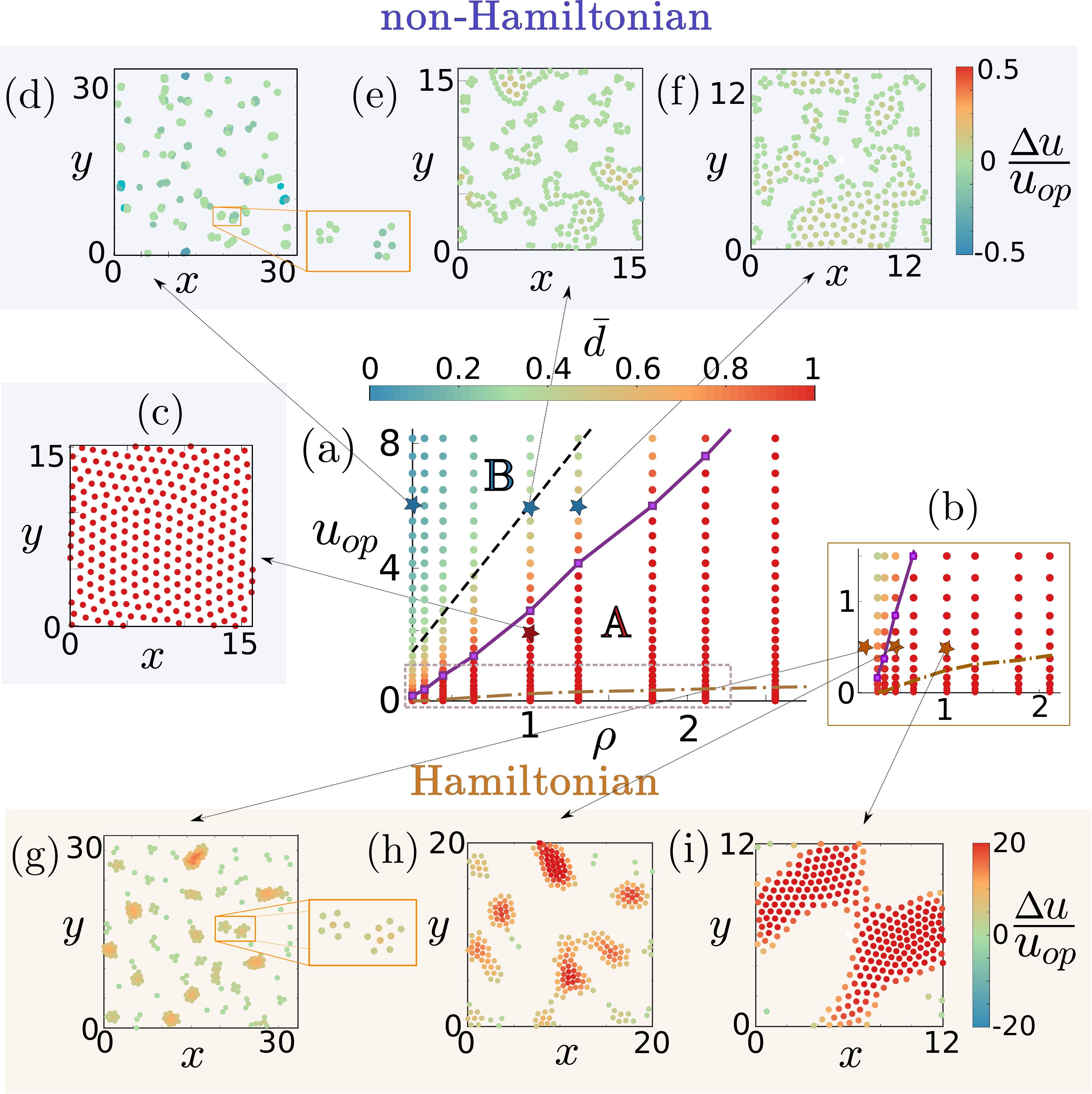}
	\end{center}
	\caption{\label{ph_d1}  Phase diagram: (a) The mean nearest neighbour (NN) distance  scaled by density,  $\bar{d} \rho^{1/2}$, depicted by colour, as a function of the number density $\rho$ and the comfortable field value $u_{op}$ for 256 particles confined in a 2D simulation box with periodic boundary conditions (PBCs). Two distinct regimes exist: A in which $\bar{d} \rho^{1/2}=1$ and the system is in the uniform phase and B with  $\bar{d} \rho^{1/2}<1$ and the system in a patterned phase. The purple line marks the transition from A to B for our non-Hamiltonian model whereas the brown dashed dotted line marks the same transition for the Hamiltonian case. The black dashed line is the prediction of the mean-field theory. Subfigure (b) provides a zoom of (a)  in the region of small $u_{op}$ values. ((c)-(i)) Configurations of 256 particles at $t=20$, with the color indicating the value of the relative field $(u-u_{op})/u_{op}$ of each  particle.
		Subfigures (c)-(f) are for the non-Hamiltonian case with (c) $u_{op}=1.96$, $\rho=1$, (d) $u_{op}=6$, $\rho=0.25$, (e) $u_{op}=6$, $\rho=1$ and (f) $u_{op}=6$, $\rho=1.3$. Subfigures (g)-(i) are for the Hamiltonian case with $u_{op}=0.49$ and
		(g) $\rho=0.11$, (h)  $\rho=0.44$ and (i) $\rho=1$.}
\end{figure}

Except for the density $\rho$, our model possesses a further control parameter, namely the optimal field value $u_{op}$. This controls essentially the optimal separation of the particles  $\bar{d}^*$, which  decreases with increasing $u_{op}$.  Thus the critical density $\rho^*$ for the transition from the uniform to a patterned state  is expected to increase with $u_{op}$. This argument is substantiated by continuum mean field calculations,
which predict \footnote{see the Methods section} that the uniform state loses its stability at $\rho^*\approx u_{op}/2\pi$.

In order to get a deeper insight into the phase behavior of our system we have monitored its long-time structural evolution in  2D simulations of overdamped dynamics  governed by the equations
\begin{equation}
\dot{\vec{r}}_i= \frac{1}{\gamma}\vec{F}_i +\gv{\eta}_i
\end{equation}
where $\gamma$  is the Stokes drag coefficient, the term $\gv{\eta}_i$
stands for a zero-mean Gaussian noise with $\avg{\gv{\eta}_i(t)\gv{\eta}_j(t+\tau)}=2 D_t \delta_{ij} \delta(\tau) \vec{I}$, and $\vec{I}$ denotes the unit matrix. Our simulations are performed
for different densities $\rho$, different $u_{op}$ values and a random initial condition. Our results are summarized in Fig. \ref{ph_d1} for the case of our full non-Hamiltonian model (Eq. \ref{full1}, Fig. \ref{ph_d1} (a),(c)-(f)) and for the Hamiltonian approximation $\vec{F}_i\approx\vec{F}_{i,p}$ (Eq. \ref{pair_int},  Fig. \ref{ph_d1} (b),(g)-(i)). For both cases (Fig. \ref{ph_d1} (a),(b)) we find   two phases, A and B. In the A phase  a uniform hexagonal crystal is formed with an inter-particle spacing which in general differs from the optimal value $\bar{d}^*$ and is dictated by the density $\rho$ (Fig. \ref{ph_d1} (c)). The B phase  is a patterned state, where clusters with a complex and aperiodic structure emerge. Remarkably, these self-organized structures allow most individuals in the system to be separated at an optimal distance from adjacent particles (Fig. \ref{ph_d1} (d)-(f)).  Furthermore, in the B phase  the clusters tend to become smaller in size and more in number with decreasing density (Fig. \ref{ph_d1} (f)-(d) and (i)-(g)), since the distance between the particles in their initial random configuration is overall increased, causing the respective interactions to be weaker.  We remark here that the solutions of the corresponding mean-field equations display a similar behaviour (see SI).

The critical density $\rho^*$ for the non-Hamiltonian case seems to scale linearly with $u_{op}$, according to the mean-field prediction,  but there is some quantitative deviation, due to the finiteness of the simulation system. 
Within the Hamiltonian approximation, $\rho^*$ occurs at very low $u_{op}$ values (Fig. \ref{ph_d1} (b)), since the particles' attraction is much stronger if the triplet interactions are absent (Fig. \ref{tri_cha1} (b),(e)). 
Note that importantly  only in the non-Hamiltonian case,
the particles accumulate in the outer part of the cluster (Fig. \ref{ph_d1} (d)-(f)), while being much more homogeneously distributed inside  the clusters for  the Hamiltonian one (Fig. \ref{ph_d1} (g)-(i)). Thus three-body interactions allow a much larger fraction of particles to reach the optimal field state, as we shall discuss in more detail below.
Notably, a similar behavior is exhibited by the mean-field model, featuring solutions with density spikes at the patterns' boundaries. This is in contrast to the volume-filling model which, similarly to the Hamiltonian approximation, displays  patterns of  density plateaus (see SI). 

\begin{center}
	{\textbf{B. Cluster structure}}
\end{center}

\begin{figure}[htbp]
	\begin{center}
		\includegraphics[width=0.6\columnwidth]{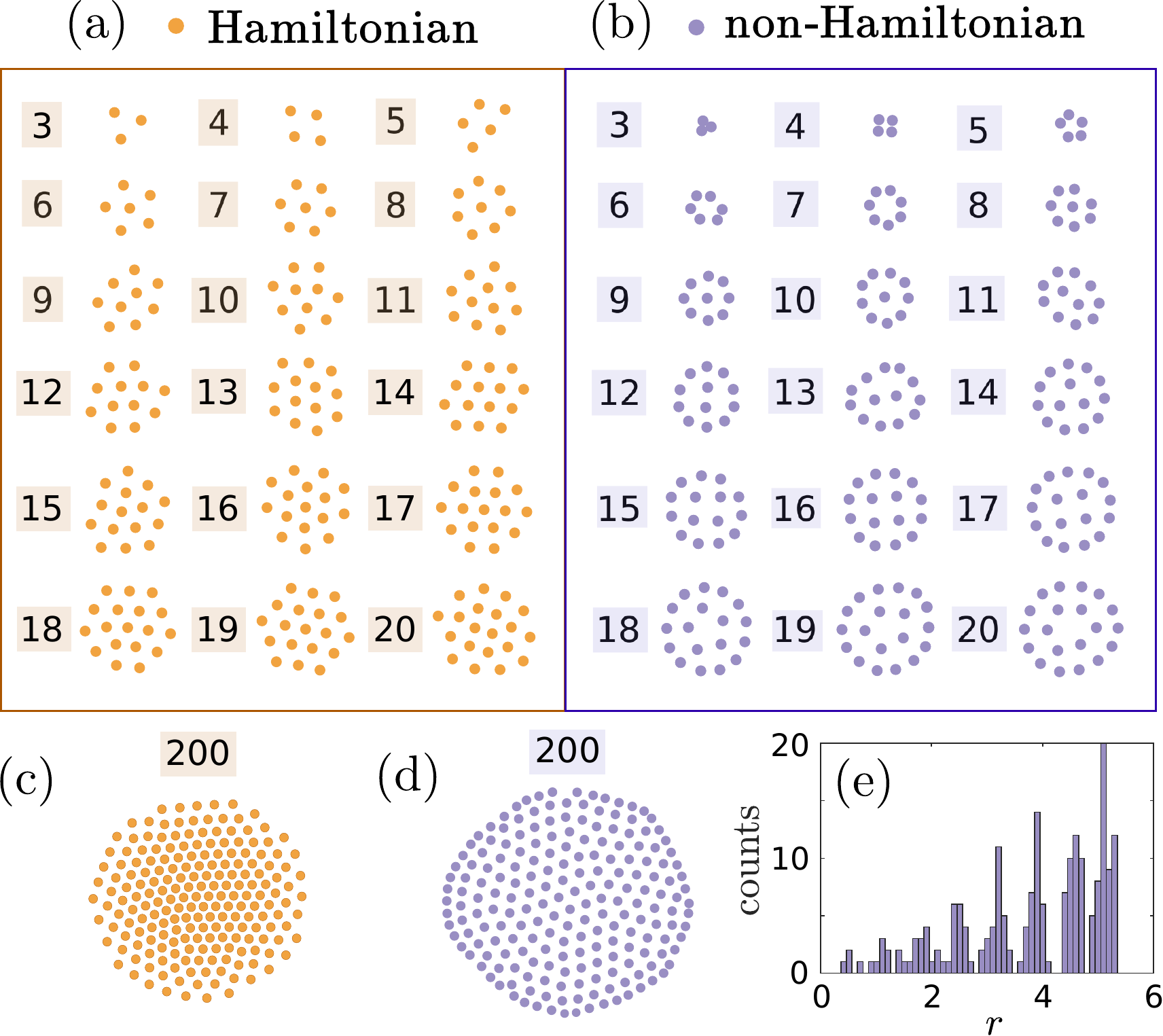}
	\end{center}
	\caption{\label{trip1}  Cluster structure: ((a),(b)) Indicative cluster structures of $N=3-20$ particles in 2D for (a) the Hamiltonian and (b) the  non-Hamiltonian case.
		((c),(d)) $200$-particle clusters in 2D.
		(e) Histogram of the particle distances $r$  from the $200$-particle cluster centre 
		for the non-Hamiltonian model. 
		For all the Hamiltonian cases,
		depicted here with the orange colour, $u_{op}=0.5$, whereas for the non-Hamiltonian ones, depicted with purple, $u_{op}=6$. Note that for all the equilibrium clusters of this figure the particles were initially positioned on a square grid. }
\end{figure}

The difference between the cluster formation in the Hamiltonian and in the non-Hamiltonian case is highlighted by investigating the equilibrium clusters formed in the two cases for a small number of particles $N=3-20$ (Fig. \ref{trip1} (a),(b)). While in the Hamiltonian case the particles form the expected  closely packed clusters (Fig. \ref{trip1} (a),(c)), in the non-Hamiltonian one  they organize in different shells (Fig. \ref{trip1} (b)) with the outer shell accommodating  most particles. This shell structure persists even for higher particle numbers (Fig. \ref{trip1} (d)) and is nicely captured by the radial distribution function (Fig. \ref{trip1} (e)).

\begin{center}
	{\textbf{C. Optimal hypersurface manifolds}}
\end{center}

In hindsight, in our $N$ particle non-Hamiltonian model, the force equilibrium condition for each particle $i$, $\vec{F}_i=\vec{0}$,  can be satisfied, according to Eq. \ref{full1},
by   $\nabla_i u(\vec{r}_i)=\vec{0}$ or $u(\vec{r}_i)=u_{op}$. Thus, except 
for the standard vanishing of a 2D gradient, an equilibrium of particle $i$ is also achieved when the total scalar field at its position  $\vec{r}_i$ is optimal. These two possibilities are present for each of the $N$ particles,
giving rise to a plethora of different equilibria.

More insight into the differences of these equilibria can be provided by regarding the two extreme cases, i.e. the one with
$\nabla_i u(\vec{r}_i)=\vec{0}$ for all $i$ and the other with $u(\vec{r}_i)=u_{op}$  for all $i$. The first case has the form of a standard minimization problem with $V_0=\frac{1}{2}\sum_{k}u(\vec{r}_k)$ in the role of the total potential. This results typically in $2N$ constraints for $2N$ degrees of freedom, from which 3 are redundant due to the central nature of the potential. The corresponding equilibrium is stable with 3 zero modes: the two global translations and the global rotation.  

Oppositely, the case $u(\vec{r}_i)=u_{op}$ provides a scalar condition for each particle and thus yields $N$
constraints for $2N$ degrees of freedom. Since $N$  degrees of freedom are free to move, the respective stable equilibria will possess $N$ zero modes, along which the system can freely deform, resulting in a very high degree of floppiness. 
Remarkably, the clusters occurring in simulations for long times (Fig. \ref{trip1} (b),(d))  are typically a combination of the above cases with some particles, usually the inner ones, satisfying the zero-gradient and others (the outer particles) the optimality condition. Importantly, there is always a tendency of the particles to accumulate on the periphery of the clusters (Fig. \ref{ph_d1} (d)-(f)), satisfying $u(\vec{r}_i)=u_{op}$, and thus generating  a large number of zero modes.

\begin{figure*}[htbp]
	\begin{center}
		\includegraphics[width=17.8 cm]{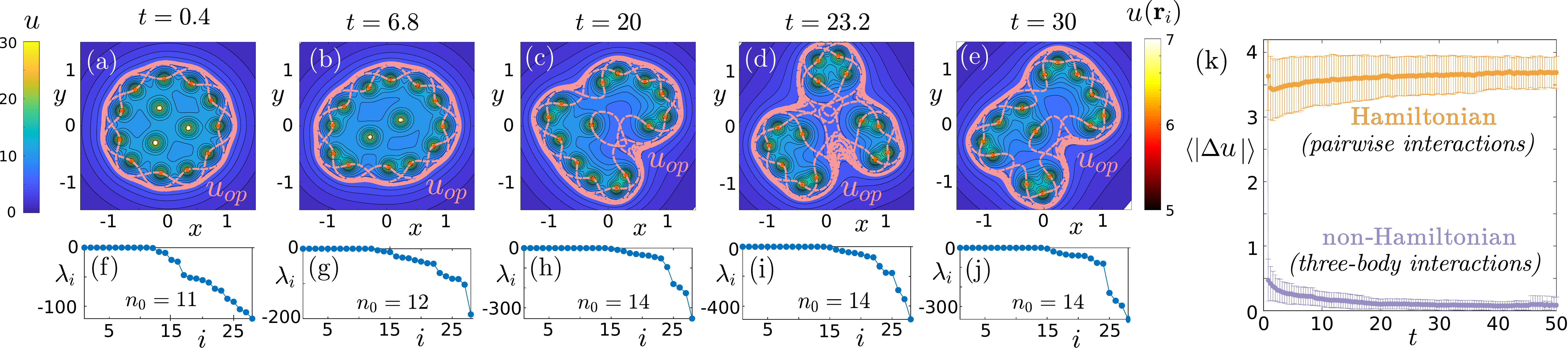}
	\end{center}
	\caption{\label{co1} Collective self-optimization: (a)-(e) Contour plot of the total  field ${u}(x,y)$ (Eq. \ref{eq_temp})
		for the equilibrium positions $\{\vec{r}_i^{(0)}\}$ of 14 particles in our non-Hamiltonian model. The presented equilibria are  realized
		at different time instants $t$ during the system's  dynamics.  For better visualization we fix the center of mass of the clusters at (0,0). The colour of each particle $i$ is indicative of its field value $u\left(\vec{r}_i\right)$. The dashed pink lines depict the set of points which satisfy the condition $u\left(\vec{r}_i\right)=u_{op}$, for each particle $i$.  (f)-(j) Sorted eigenvalues $\lambda_i$ of the Jacobian of the system around the equilibrium positions $\{\vec{r}_i^{(0)}\}$, shown at subfigures (a)-(e). (k) Time evolution of the mean deviation from the optimal state $\avg{\abs{\Delta u}}$, averaged over 50 realizations, for the Hamiltonian (orange) and the non-Hamiltonian (purple) model with $N=14$. The errors are estimated by the corresponding standard deviations. All the simulations were performed for a comfortable field value $u_{op}=6$ for the non-Hamiltonian and $u_{op}=0.5$ for the Hamiltonian model and  for a translational diffusion coefficient $D_t= 0.003$, while the particles were constrained  in a 2D simulation box.
	}
\end{figure*}

We exemplary visualize this for $N=14$ particles, where  we can see  that soon after their initialization in a square grid, the field value $u$ of the outermost 11 particles is the optimal one, $u_{op}$, while the inner 3 particles lie at larger field values (Fig. \ref{co1} (a)). According to our above discussion this leads to a large number, $n_0=11$, of zero eigenvalues of the Jacobian (Fig. \ref{co1} (f)), implying a free  distortion along the corresponding eigenvectors. 

The floppiness of this cluster leads with even a slight noise to its continuous deformation in time. Importantly, during this dynamics many other different equilibria are visited (Fig. \ref{co1} (b)-(e)) which also possess a very large number, $n_0$, of zero eigenvalues $\lambda_i$ of the Jacobian (Fig. \ref{co1} (g)-(j)). Surprisingly and in direct contrast to the Hamiltonian case where the optimal state cannot be reached, for the non-Hamiltonian system there is overall a tendency  to minimize the mean deviation from the optimal state, measured by $\avg{\abs{\Delta u}}=\frac{1}{N} \sum_j \abs{ u(\vec{r}_j)-u_{op}}$,  in the course of time (Fig. \ref{co1} (k)).  This demonstrates the importance of three-body forces for the collective self-optimization in the system.
An optimal state for each particle is indeed reached for quite long times (Fig. \ref{co1} (c)), but even then, the cluster continues deforming to different optimal states (Fig. \ref{co1} (d),(e)), due to its high degree of floppiness.  Notably, the intricacy of this dynamical behaviour, cannot be captured within a mean-field model (see SI), since it involves a substantial restructuring of particles positions.

\begin{center}
	{\textbf{IV. DISCUSSION}}
\end{center}

 We have introduced a simple model of agents communicating via a shared field which they produce or consume, and aiming to achieve a certain optimal field value.   The proposed communication rules ensure that each agent approaches its individual optimum as time evolves.  Effective three-body interactions, stemming from the very existence of  an optimal field value, play a crucial role here. These interactions render the system non-Hamiltonian and highly flexible, 
providing thus the particles with numerous pathways to reach collectively their desired state. The latter turns out to be highly degenerate, causing continuous deformations of the particles' configurations in time, a fact that resembles the
dynamical complexity of living clusters observed in nature.

In contrast to game-theoretical agent models used often in socio-economical studies \cite{Farooqui2016,Fletcher2008, Stewart2014}, our model does not rely on an appropriate choice of a benefit and a cost function, determining the pay-off  of each individual. Instead, it puts the emphasis on the physical modelling of the  communication between the agents, via their (produced or consumed) shared field,  as established in the framework of chemotaxis. In this way, the simple rule of separately steering  their motion towards their desired  field value creates complex interdependencies between the agents, which are embodied in the emerging three-body interactions. The latter act as  an "invisible hand", assisting the group to reach its collective optimum.  This observation may inspire future studies, in the framework of mean-field games and optimal control theory, to assess the persistence and efficiency of the employed communication scheme in a game theoretical scenario, similarly to \cite{Yin2014,Grover2018,Tania2012}. For agents that consume a finite resource, this chemotactic scheme can lead to a social dilemma situation \cite{MacLean2008} where the impact of selfishness and cooperation can be explored \cite{Fletcher2008,Levin2014}.

We point out that the dimensionality of the system affects greatly its ability  to find an optimal state. The 2D space studied here offers enough flexibility for restructuring, allowing the particles to reach collectively an optimal state -- this is not the case for 1D systems where the particles' motion is too restricted. Furthermore, an important assumption of our model is that the optimal field value is the same for all individuals,  a fact that  introduces a high degree of symmetry in the system and allows ultimately for an alignment of their interests. It would be interesting therefore to investigate in future studies, to what degree a heterogeneity in the optimal field  affects the collective self-optimization in such systems. Notably, for heterogeneous populations, elaborate schemes, such as a division-of labour into follower and leader roles \cite{Pais2014} are usually required to promote  the individuals' cooperation.

Regarding the relation of our model to previously studied chemotactic models,  we note that out of many important variations of the original Patlak-Keller-Segel model \cite{Hillen2009}, only some of them allow for a Lagrangian particle-based formulation as  in this work. A prominent example here is the discussed volume-filling model, which (since it features an optimal density) is inherently linked to an Eulerian description. On the other hand, the Lagrangian approach is particularly useful, since it does not only allow us to explore some distinguishing properties of chemotaxis \cite{Calvez2016,Stevens2000,Newman2004,Romanczuk2008,Tyson1999}, but also 
enhances the link to the growing research field of active matter \cite{Liebchen2019,Bechinger2016}. This generates new insights -- for instance, we expect that the optimization
principles proposed here will
be of relevance for other realizations of active matter in complex environments  \cite{Bechinger2016}. These include
optimal search strategies for animals looking for food or mates, or
for rescuers in disaster zones \cite{Volpe2017}. Moreover, the Lagrangian approach enables an explicit investigation of structural dynamics, including dynamical changes of the particles' neighbourhood, which  play a decisive role in collective behavior \cite{Klamser2021}. 

Further connections of our model to collective behavior can be drawn by considering the repulsion-attraction character of non-local interactions, essentially resembling models often used to explore the collective properties of swarming systems \cite{Chuang2007,Leverentz2009}. While the phase behaviour of our model is similar to that of passive particles interacting with a generalized Morse potential \cite{Orsogna2006}, the formed clusters display a different structure and dynamics, which is a result of non-Hamiltonian three-body interactions. It is thus plausible that an extension of our model, including an explicit self-propulsion and possibly also a dissipation of the particles' motion, could reveal an intriguing morphology and evolution of swarming states.

Finally, we believe that, due to its straightforward formulation,  our model  can draw a connection to the field of artificial swarm intelligence.  In the same sense, the minimal yet intelligent communication rules employed here could be realized with state-of-the-art techniques for programmable interactions in synthetic active particles  \cite{Bechinger2020,Vutukuri2020}.

\appendix
\begin{center}
	{\textbf{METHODS}}
\end{center}

\begin{center}
	{\textbf{A. Mean field theory}}
\end{center}

In the  continuum description of our system's dynamics in terms of the scalar field  and the particle density  we need to distinguish between the 3D quantities,  $u_{3D}\left(\vec{r},t\right)$ and $\rho_{3D}\left(\vec{r},t\right)$, and the corresponding 2D ones, denoted  by $u(x,y,t)$ and $\rho(x,y,t)$, respectively. 

As we have discussed, the evolution of the field is assumed to be governed
by the diffusion equation \ref{dif_eq}. In units of $x_u=\kappa^{-1}$, $u_u= A \kappa$ and $t_u=\gamma/(\lambda A \kappa^4)$ for the length, the field and the time this reads
\begin{equation}
\pd{u_{3D}}{t}=\frac{1}{4\pi} \nabla^2  u_{3D}-\frac{1}{4\pi}u_{3D} +\rho_{3D}, \label{dif_eq3}
\end{equation}
Since in our model the particles are constrained to move  on the 2D $xy$ plane, we have 
$\rho_{3D}(\vec{r},t)=\rho(x,y,t)\delta(z)$. The 2D number density  $\rho(x,y,t)$ follows  the
equation of motion
\begin{equation}
\pd{\rho}{t}=\nabla_{||} \cdot \left[\rho\left(u-u_{op}\right) \nabla u\right]+ D_t \nabla_{||}^2 \rho ,\label{dif_eq2}
\end{equation} 
where  $u(x,y,t)=u_{3D}(x,y,0,t)$, $u_{op}$ is the optimal field value,  $\nabla_{||}=\left(\frac{\partial}{\partial x},\frac{\partial}{\partial y}\right)$ denotes the 2D del operator and $D_t$ stands for the translational diffusion coefficient.

The static solution of the coupled equations  \ref{dif_eq3} and \ref{dif_eq2}, corresponding to a constant density $\bar{\rho}$ in the $xy$ plane,  reads
\begin{equation}
\rho= \bar{\rho},~u_{3D}={2\pi \bar{\rho}}e^{-\abs{z}},\label{sol_fiel1}
\end{equation}
This solution  is representative of the solid/uniform phase (A) of our system (see also Fig. \ref{ph_d1}).

Using the ansatz $\rho= \bar{\rho}+r\exp\left(iqx-\lambda t\right)$
and  $u_{3D}= {2\pi \bar{\rho}}e^{-\abs{z}}+f(z)\exp\left(iqx-\lambda t\right)$ we perform a linear stability analysis around the homogeneous equilibrium (Eq. \ref{sol_fiel1}). This yields the following expression for the eigenvalues $\lambda$:
\begin{equation}
\left(\lambda-D_tq^2\right)\left(1+q^2-4\pi\lambda\right)^{1/2}=2\pi \bar{\rho}\left(2\pi \bar{\rho}-u_{op}\right)q^2.
\end{equation}
In the limit $q \rightarrow 0$ the lowest eigenvalue can be approximated by

\begin{equation}
\lambda_{min}\approx\left[D_t+2\pi \bar{\rho}\left(2\pi \bar{\rho}-u_{op}\right)\right]q^2. \label{eig2}
\end{equation}
Stability is then provided if 
$\lambda_{min} \geq 0$, yielding 
\begin{equation}
\bar{\rho}\geq\frac{u_{op}+\sqrt{u_{op}^2-4 D_t}}{4 \pi }. \label{stab11}
\end{equation}
As a result, for the case of zero noise ($D_t=0$) the homogeneous solid phase is stable if $\bar{\rho} \geq \frac{u_{op}}{2\pi}$ and unstable in the opposite case.
We note  that the stability analysis on the level of mean-field theory, presented here, is  equivalent to that of the volume filling model, with $2\pi \bar{\rho}$ changed to $\rho$ and $u_{op}$ to $\rho_{op}$. More information on the comparison between the two models is provided at the SI.

\begin{center}
	{\textbf{B. Simulations}}
\end{center}

The numerical results presented  in this article were 
obtained by 2D Brownian dynamics simulations based on the
overdamped equations of motion

\begin{equation}
\dot{\vec{r}}_i= \frac{1}{\gamma}\vec{F}_i +\gv{\eta}_i
\end{equation}
where $\vec{r}_i$ is the position of particle $i$ and $\gamma$  denotes the friction coefficient. The term $\gv{\eta}_i$
stands for a zero-mean Gaussian noise with $\avg{\gv{\eta}_i(t)\gv{\eta}_j(t+\tau)}=2 D_t \delta_{ij} \delta(\tau) \vec{I}$, where  $\vec{I}$ is the unit matrix. Similar equations of motion have been used for active particles in a different context, see e.g. \cite{Romanchuk2012,Volpe2014,Reichhardt2018, Klapp2020}.
Here, the force
$\vec{F}_i$ is provided by Eq. \ref{full1} for our non-Hamiltonian model and by Eq. \ref{pair_int} for the Hamiltonian approximation. In all cases we have used periodic boundary conditions (PBCs) and we have chosen the  units $x_u=\kappa^{-1}$, $u_u= A \kappa$ and $t_u=\gamma/(\lambda A \kappa^4)$ to measure  the length, the field and the time, respectively. In these units we have
that $\gamma=1$, $\kappa=1$, $\lambda=1$ and the value of the integration step we used reads $dt=10^{-4}$. Unless stated otherwise the simulations were performed with $D_t=3 \times 10^{-5}$.

For the phase diagram of Fig. \ref{ph_d1} the density was tuned by changing accordingly the square box size $L$. Each point of Fig. \ref{ph_d1} (a) is the result of a single run for $N=256$ particles,
for the same random initial configuration, propagated up to a final time $t=20$. For the clusters of Fig. \ref{trip1}  the particles were initialized on a square grid of side length $a=0.3$ inside a square box of size $L=10$ and were evolved until a time $t=10$. In order to find the equilibrium structures, satisfying $\vec{F}_i=0$, we have used these resulting configurations as an initial guess for a Newton root-finding method.

\begin{center}
{ \textbf{ACKNOWLEDGEMENTS}}
\end{center}
This work is supported by the German Research Foundation through grants
LO 418/23-1 and IV 20/3-1.

\begin{center}
	{ \textbf{AUTHOR CONTRIBUTIONS}}
\end{center}
B.L. and H.L. designed research; A.V.Z. performed research; A.V.Z. analyzed data; and A.V.Z., B.L., A.V.I., and H.L. wrote the paper.

\begin{center}
	{ \textbf{DATA AVAILABILITY}}
\end{center}
Results of numerical simulations and corresponding code have been deposited in GitHub (10.5281/zenodo.5223622). All other study data are included in the article and/or SI Appendix.

\bibliography{pen-sample}
\end{document}